\documentclass[RNAAS]{aastex631}
\usepackage{multirow}
\usepackage{amsmath}
\usepackage{footmisc}

\shorttitle{Search for extended sources  in  Chandra ACIS images }
\shortauthors{Volkov \& Kargaltsev}

\begin{document}

\title{Search for extended sources in the images from Chandra  X-ray Observatory Advanced CCD Imaging Spectrometer }

\correspondingauthor{Oleg Kargaltsev}
\email{kargaltsev@gwu.edu}

\author{Igor Volkov}
\altaffiliation{}
\affiliation{The George Washington University}

\author{Oleg Kargaltsev}
\altaffiliation{}
\affiliation{The George Washington University}

\keywords{methods: statistical; techniques: image processing;  Astrophysics - Instrumentation and Methods for Astrophysics; Astrophysics - High Energy Astrophysical Phenomena}

\section{Abstract}

We present a convenient tool (ChaSES) which allows to search for extended  structures in Chandra  X-ray Observatory Advanced CCD Imaging Spectrometer (ACIS) images. The tool  relies on DBSCAN clustering algorithm to detect regions with overdensity of photons compared to the background. 
 Here we describe the design and functionality of the tool which we make publicly available on GitHub. We also provide online   extensive   examples of its applications to the real data.

\section{Background} 

Chandra X-ray Observatory (CXO) Advanced CCD Imaging Spectrometer (ACIS; \citealt{2003SPIE.4851...28G}) has taken thousands of images with unprecedented sub-arcsecond angular resolution and very low  background. Therefore, even shallow Chandra images provide an  opportunity 
 to look for faint extended sources of X-ray emission (such as supernova remnants, pulsar-wind and magnetar-wind nebulae, galaxy clusters, shocks driven by massive stars or star clusters, planetary nebulae, etc.).  Although, Chandra is well known for  spectacular images of bright extended sources, there are no convenient tools that allow to look for fainter extended structures that may be serendipitously imaged while observing other targets. 
  Given the large volume of data collected by CXO, 
  such a search would require fast, efficient, and robust structure-finding algorithm. The standard CIAO source-detection tools\footnote{These are the wavedetect, celldetect, and vtpdetect tools described in https://cxc.harvard.edu/ciao/download/doc/detect\_manual/} available in CIAO\footnote{Software package developed by CXC to analyze Chandra data: http://cxc.harvard.edu/ciao/} are mostly geared toward point source detection and, hence, do not provide accurate characterization of significantly extended (compared to the {\sl CXO} PSF) sources.
  Therefore, after extensive comparison of various algorithms, we opted to use the well-known and well-tested DBSCAN clustering algorithm \citep{10.5555/3001460.3001507} available within the {\tt scikit-learn} Python library\footnote{https://scikit-learn.org/}.

 \section{Under the Hood: DENSITY-BASED CLUSTERING with DBSCAN}
 
At the core of the  ChaSES Tool is the scikit-learn's implementation of the DBSCAN clustering algorithm\footnote{https://scikit-learn.org/stable/modules/generated/sklearn.cluster.DBSCAN.html}. This is a  density-based  algorithm capable of finding cluster of arbitrary shape and variable density in the presence of noise.   A cluster is a set of core points (such that there exist $min\_samples$ other points within a distance of $\epsilon$) which is built by recursively taking a core point, finding all of its neighbors that are core samples, finding all of their neighbors that are core points, and so on. It also includes a set of non-core boundary  points  (neighbors of a core point in the cluster but are not the core points themselves). Higher $min\_samples$ or lower $\epsilon$ correspond to higher density. The optimal vales of $min\_samples$ or lower $\epsilon$ depend on the dataset and are often found by the trial-and-error method. There are no universally optimal values. However, we found the values that are appropriate for most ACIS  datasets and set them as defaults in the ChaSES GUI (see below). Users are encouraged to vary these parameters in the vicinity of these values.

   \section{Pre-processing, Graphical User Interface (GUI), and Output}

The ChaSES tool\footnote{Available at https://github.com/ivv101/ChaSES} can be run in the user's web browser 
or in the a Python/Jupyter  notebook. ChaSES allows users to analyze any observation existing in the CXO archive which can be specified by the unique observation ID (ObsID). The analysis consists of two parts: (1) creation of energy-filtered  event list with  point sources being removed and (2) search for extended structures by running DBSCAN with  the results shown graphically and in  tabular form. The  regions corresponding to clusters can be  exported in SAO DS9 format.

Since removal of point sources requires user to have CIAO installed and also computationally expensive, we pre-computed event lists that have point source removed\footnote{Note that the removal can be imperfect but it does, on average, help to increase sensitivity to extended sources.} for 
1,042 ACIS observations and put those files onto Hugging Face\footnote{https://huggingface.co/datasets/oyk100/Chandra-ACIS-clusters-data} (HF) website\footnote{Due to the large volume of data these files could not be placed  on GitHub.}.  Users can directly download those files from our ChaSES 
GUI\footnote{ Implemented with Bokeh Python library: https://docs.bokeh.org/ } by using the ``pre-computed'' button and selecting any of the available ObsIDs. For each ObsID the photons in the ACIS event list are filtered to the 0.5-8 keV energy range. (The  original event list with point sources is also  included.)  
On  GitHub  we provide the script\footnote{See the corresponding CIAO thread  https://cxc.cfa.harvard.edu/ciao/threads/diffuse\_emission/}.)  that can be used  to remove point sources from any ACIS observation, when the corresponding ObsID is not present among the pre-processed datasets in our HF repository\footnote{Note, this step requires installing CIAO and downloading the entire set of data products associated with a particular ObsID. This can be done with the help of CIAO's {\tt download\_chandra\_obsid script.} }.

Once the ObsID is selected in the ChaSES GUI, the user can  select the  CCD  needs to choose the corresponding drop down menu (``ccd'') since the cluster search is performed per single CCD (to avoid the interference with the gaps separating CCDs). The ``holes'' button allow one to switch between the original event files and files where the point sources have been removed. When pressed, the ``n\_max'' button randomly under-samples the data by capping the number of photons at 20,000. This is helpful when the the dataset is large and hence computations can take long time. (De-pressing the button will cause ChaSES run on the entire dataset.) In addition to the above, the user can change 

\begin{itemize}
    \item the number of pixels in the image: $nbins$, and 
    \item the DBSCAN algorithm parameters: $\epsilon$ (``eps'' in the GUI) and  $min\_samples$.
\end{itemize}

The $\epsilon$ ($eps$) and  $min\_samples$ are the two main parameters of the DBSCAN clustering algorithm. By default they are set to the values that should be closed to the optimal ones (see above). The cluster search is initiated by pressing the  ``Apply'' button in GUI. Once the clusters are found by DBSCAN, the background is determined by calculating the average number of photons per image pixel after excluding the regions associated with the clusters. Therefore, it is dependent on the choice of  $nbins$,  $eps$, and  $min\_samples$ (which should be sensible).  The average background value, after multiplying by it by the cluster area, is used to calculate the chance occurrence probability ($P_{\rm chance}$) of the observed number of photons in the cluster (which is the extended source significance) according to Poisson distribution.

The output of ChaSES consists of the visualization of detected clusters on top of the image (with cluster regions numbered and overplotted with  different colors) and the tables with the properties of detected clusters (extended sources).   These properties are the silhouette score (one of the metrics used to characterize the quality of cluster), area (as a fraction of the total image area), number of net counts ($n-n_{\rm bkg}$), detection significance, $S$ (``signif.'' in GUI; in units of Gaussian standard deviation\footnote{$1-P_{\rm chance}=\left[\int_{-\infty}^{S } (2\pi)^{-1/2}\exp^{-x^2/2}dx\right]^{A/A_c}$, where $A$ and $A_c$ are the CCD and cluster's areas, respectively.}), and the position of cluster center of mass in the physical ($x,y$) and celestial coordinates ($R.A., Decl.$) coordinates. The table can be filtered by detection significance using the $min sigma$ slider located below the image.
The remaining adjustable parameters only affect the appearance of the image that is shown but do not affect how the detection is performed.

\section{Summary}
We developed a Python-based tool with a convenient GUI which streamlines the detection and characterization of extended sources in {\sl CXO} ACIS data. It can also be easily adopted for use with  imaging data from other X-ray telescopes.

\begin{figure}[h!]
\begin{center}
\includegraphics[width=16cm,angle=0]{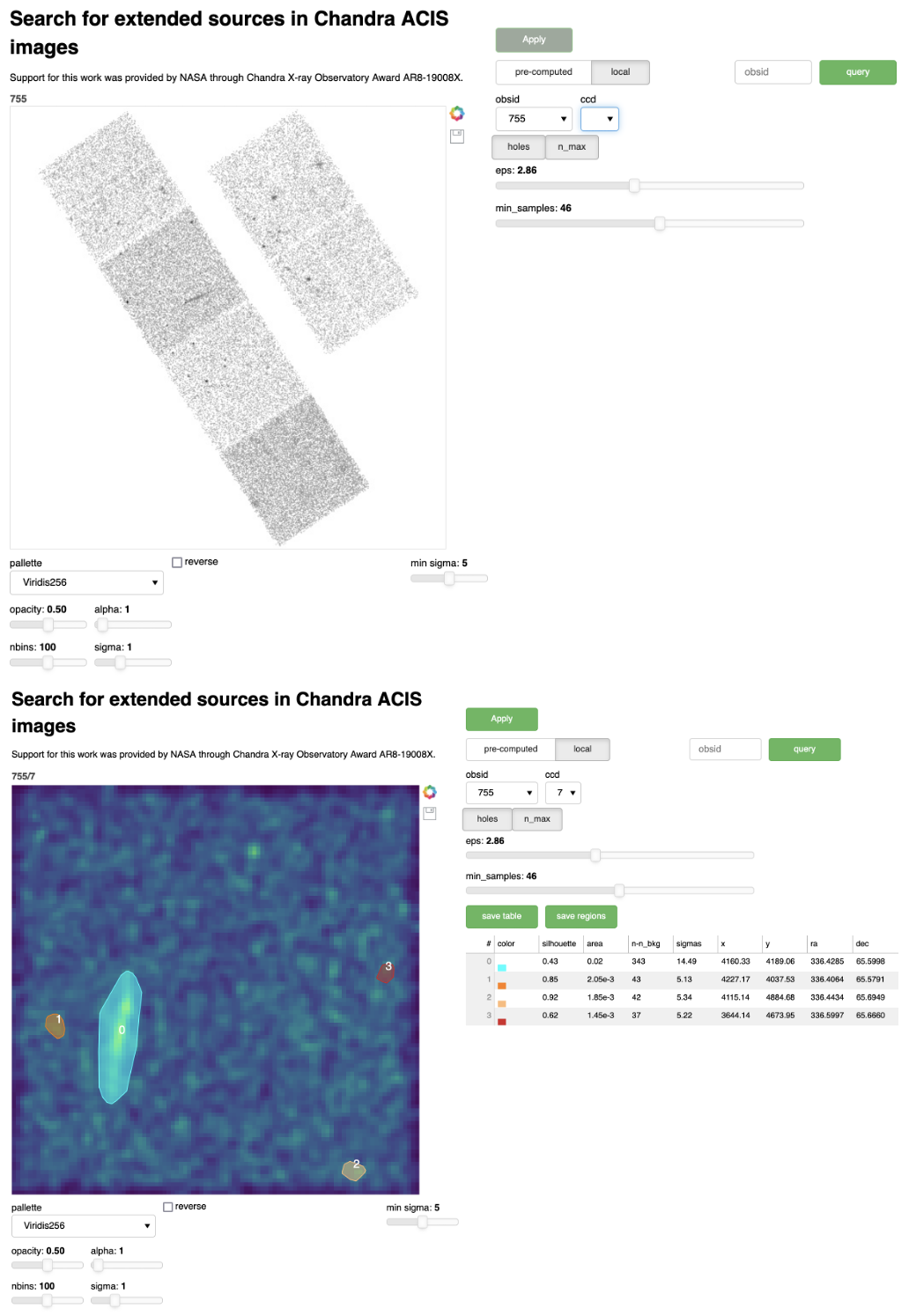}
\caption{
ChaSES GUI. The the top panel (note, the ``ccd'' selection is empty) shows the image from CXO ObsID 755 which includes all activated ACIS CCDs. North is up, East is to the right. The bottom panel shows ccd=7 image  which zooms on the nebula  associated with  PSR B2224+65. (Note that this image is oriented differently.)  }  
\label{fig:1}
\end{center}
\end{figure}

\section{Acknowledgement}

Support  for  this  work  was  provided  by NASA through Chandra X-ray Observatory  Award AR8-19008X.

\newpage

\bibliography{ms}{}
\bibliographystyle{aasjournal}

\end{document}